\documentstyle[sprocl]{article}

\newfont{\prova}{cmsy10 scaled 1200}
\newcommand{\text}[1]{\mbox{#1}}
\input{psfig}

\def\be{\begin{equation}}
\def\ee{\end{equation}}
\def\bea{\begin{eqnarray}}
\def\eea{\end{eqnarray}}

\begin{document}

\title{CHERN--SIMONS FIELD THEORIES IN THE COULOMB GAUGE}

\author{\underline{FRANCO FERRARI}$^*$ and IGNAZIO LAZZIZZERA $^{**}$}

\address{$^*$LPTHE, Universit\'e Paris VI and VII, 4 Place Jussieu,
\\ F-75252 Paris CEDEX 05, FRANCE\\E-mail: fferrari@lpthe.jussieu.fr} 
\thanks{%
$^b$Laboratoire associ\'e No. 280 au CNRS}

\address{$^{**}$Dipartimento di Fisica, Universit\`a degli Studi di Trento,
Via Sommarive, 14\\
I-38050 Trento, Italy\\E-mail: lazi@science.unitn.it}


\maketitle\abstracts{ In this talk some recent results in the
quantization of Chern-Simons field theories in the Coulomb gauge
will be presented.
In the first part,  the consistency of the
Chern--Simons field theories in this gauge is proven
using the Dirac's canonical formalism for constrained
systems.
Despite the presence of non-trivial self-interactions in the
gauge fixed functional, it will be shown that
the commutation relations between the fields
are trivial at any perturbative order
in the absence of couplings with matter fields.
If these couplings are present, instead, the commutation relations become
rather
involved, but it is still possible to study their main properties and to show
that they vanish at the tree level.
In the second part of the talk the perturbative aspects of
Chern--Simons field theories in the Coulomb
gauge will be analysed. In particular, it will be shown by explicit
computations and in a regularization independent way
that there are no radiative contributions to the
$n-$point correlation functions.
Finally the Feynman rules in the Coulomb gauge will be derived on a
three dimensional manifold with a spatial section given
by a closed and orientable Riemann surface.
}

\section{Introduction}

In the recent past, the Chern--Simons (C--S) field
theories~\cite{jao,hageno} have intensively been studied in connection
with several
physical and mathematical applications~\cite{csapps,ienkur}.
A convenient gauge fixing for these theories is provided by the Coulomb gauge.
As a matter of fact, the presence of nontrivial
interactions in the gauge fixed action allows perturbative
computations. Perturbation theory is important in the
calculations of
the so-called link invariants~\cite{witten,gmm,cotta,axelrod}
and whenever interactions are present,
because in the latter
case the C--S field theories are no longer exactly solvable.
The advantage of the Coulomb gauge in this case is that the calculations are
considerably simpler than in the covariant gauges and there are no
radiative corrections. Moreover, with respect to the axial and light
cone gauges, the Coulomb gauge can easily be imposed also on
manifolds with non-flat spatial sections, like for instance
Riemann surfaces. The absence of quantum contributions is a great
advantage on  non flat space-times, where the computation of Feynman
integrals becomes technically difficult.
Another important feature of the C--S field theories in the Coulomg
gauge is that they can be considered as two dimensional models as it
will be shown below.

Starting from the seminal works of refs.~\cite{hageno,hagent} and~\cite{jat}, 
the Coulomb gauge has been already applied in a certain number of
physical problems involving C--S based
models~\cite{ienkur,vo,devone,devtwo,bcv}, but still remains
less popular than the covariant and axial gauges.
One of the main reasons is probably the fact that there are
many
perplexities concerning the use of
this gauge fixing,  in particular in the case of the four dimensional
Yang--Mills theories~\cite{taylor,chetsa,leibbrandt,leiwil}.
Recently, also the consistency of the C--S field theories in the
Coulomb gauge has been investigated using various
techniques~\cite{ffprd,devone,cscgform,frig},
and a detailed perturbative analysis of the non-abelian case
has been done in~\cite{flpert}.

In this talk some recent results in the quantization of C--S field
theories in the Coulomb gauge will be presented following
refs.~\cite{frig,flpert,ffunp} and avoiding
technical details as much as possible.
In the first part of the talk the C--S field theories are analysed
by means of the Dirac's formalism for constrained systems.
Besides some subtleties already noticed in~\cite{linni},
the derivation of the final Dirac brackets requires in the
Coulomb gauge some care with distributions.
Moreover, the final commutation
relations (CR's) between the fields
are derived both in the case of pure and interacting C--S field theories.
With respect to the Yang--Mills field theories the CR's are rather
involved.
At a first sight, this is surprising in topological
field theories with vanishing
Hamiltonian and without degrees of freedom. However, at
least in the pure C--S field theory,
in which there are no interactions with matter fields,
we show that
this complexity is only apparent. As a matter of fact,
taking into account the Gauss law
and the Coulomb gauge fixing, the commutation
relations between the gauge fields vanish identically
at any perturbative order as expected.
In this way the Chern--Simons field theories in the Coulomb
gauge are not only perturbatively finite as has already been checked
in the covariant gauges~\cite{covgau},
but also free.
This is  not a priori evident, because in the Coulomb gauge the 
C--S functional contains non--trivial self--interaction terms.
In the interacting case it is only possible to prove that the CR's are
zero at the zeroth order approximation in perturbation theory.
At higher orders  however they are in general different from zero
and have a very complicated expression. This is probably due to the fact that
C--S field theories admit states with non-standard statistics.

In the second part of this talk
the radiative corrections of the Green functions
are computed at any loop order and it is shown in a regularization
independent way
that they vanish identically.
No regularization is needed for the ultraviolet and infrared
divergences since, remarkably, they do not appear in the amplitudes.
The vanishing of the quantum corrections is in agreement
with the triviality of the
commutation relations found using the
Dirac's canonical approach to constrained systems.
It is important to notice that the absence of any quantum
correction despite the presence of nontrivial self-interactions in the
Lagrangian
is a peculiarity of the Coulomb gauge that cannot be totally
expected from the fact that the theories under consideration
are topological, as finite renormalizations of the fields and
of the coupling constants are always possible.
For instance, in the analogous case of the covariant gauges,
only the perturbative
finiteness of the C--S amplitudes has been shown~\cite{csformal}
in a regulatization
independent way exploiting BRST techniques~\cite{ss}.
Indeed, a finite shift of the C--S coupling constant has been observed
in the Feynman gauges by various authors~\cite{shift,alr}.

Finally, the Feynman rules of the C--S field theories will be derived
also on a manifold whose spatial section is a Riemann surface of genus
$g$

\section{Canonical Quantization of the C--S field Theory in the Coulomb gauge }

\subsection{Notations}
In this Section we will use the following notations.
The Lagrangian of the pure $SU(N)$ C--S field
theory is given by
\begin{equation}
L_{CS}=\frac s{8\pi }\epsilon ^{\mu \nu \rho }
\left( A_\mu ^a\partial _\nu A_\rho
^a-\frac 13f^{abc}A_\mu ^aA_\nu ^bA_\rho ^c\right)  \label{lagrangian}
\end{equation}
where $s$ is a dimensionless coupling constant and $A_\mu ^a$ is the gauge
potential. Greek letters $\mu ,\nu ,\rho ,\ldots =0,1,2$ denote
space--time indices, while the first latin letters $a,b,c,\ldots =1,\cdots,
N^2-1$ denote color indices.
Moreover, the totally antisymmetric tensor $%
\epsilon ^{\mu \nu \rho }$ is defined by the convention $\epsilon ^{012}=1$.
The metric is given by
$g_{\mu \nu }=\enskip $diag$(1,-1,-1)$.
To derive the C--S Hamiltonian $H_{CS}$
we have to compute the canonical momenta: 
\begin{equation}
\pi ^{\mu ,a}\left( {\bf x},t\right) =\frac{\delta S_{CS}}{\delta \left(
\partial _0A_\mu \left( {\bf x},t\right) \right) }  \label{cmomdef}
\end{equation}
where  $S_{CS}=\int d^3xL_{CS}$, $t=x^0$
and ${\bf x}=\left( x^1,x^2\right) $.
A straightforward calculation shows that: 
\begin{equation}
H_{CS}=-\int d^2{\bf x}A_0^a\left( D_i^{ab}\pi ^{i,b}+\partial _i\pi
^{i,a}\right) \label{wrrr}
\end{equation}
In the above equation the following convention
has been used for the spatial components of the
covariant derivative:
$
D_\mu ^{ab}\equiv \partial _\mu \delta ^{ab}+f^{abc}A_\mu ^c$.
Finally, the nonvanishing equal time
Poisson brackets (PB) among the canonical variables read as follows: 
\[
\left\{ A_\mu ^a\left( {\bf x},t\right) ,\pi _\nu ^b\left( {\bf y},t\right)
\right\} =\delta ^{ab}g_{\mu \nu }\delta ^{(2)}\left( {\bf x}-{\bf y}\right) 
\]

\subsection{Constraints and intermediate Dirac brackets}
From eqs. (\ref{lagrangian}) and (\ref{cmomdef})
we obtain the following primary constraints:
\begin{eqnarray}
\varphi ^{0,a} &=&\pi ^{0,a}  \label{fpconstr} \\
\varphi ^{i,a} &=&\pi ^{i,a}-\frac s{8\pi }\epsilon ^{ij}A_j^a\qquad \qquad
\qquad i=1,2  \label{spconstr}
\end{eqnarray}
where $\epsilon ^{ij}$, $i,j=1,2$, is the two dimensional totally
antisymmetric tensor defined by $\epsilon ^{12}=1$.
Following the Dirac procedure for constrained systems,
the latter will be imposed in the weak sense:
$\varphi^{\mu,a}\approx 0$.
To this purpose, we
construct the
extended Hamiltonian: 
\begin{equation}
\widetilde{H}_{CS}=H_{CS}+\int \lambda _\mu ^a\varphi ^{\mu ,a}d^2{\bf x}
\label{extham}
\end{equation}
where the $\lambda _\mu ^a$'s represent the
Lagrange multipliers corresponding to the
primary constraints $\varphi^{\mu,a}$.

From the consistency conditions
$\dot \varphi
^{\mu ,a}=\left\{ \varphi ^{\mu ,a},\widetilde{H}_{CS}\right\} \approx 0$,
we obtain the secondary constraint:

\begin{equation}
\text{{\prova G}}^a=
D_i^{ab}\pi ^{i,b}+\partial _i\pi
^{i,b}\approx 0 \qquad\qquad\qquad \text{(Gauss law)} \label{glaw}
\end{equation}
and two relations which  determine the Lagrange multipliers $\lambda_1$ and
$\lambda_2$:
\begin{equation}
\frac s{4\pi }\epsilon ^{ij}\left( D_j^{ab}A_0^b-\lambda _j^a\right)\approx
0\qquad \qquad \qquad i=1,2  \label{lagdet}
\end{equation}
It is possible to see that the
consistency condition  \.{\prova G}$^a\approx 0$ does not lead to any
further independent equation.
The operators $\text{\prova G}^a$ generate the $SU(N)$ group of gauge
transformations only after eliminating the second class constraints
$\varphi _i^a\approx 0$ 
of eq. (\ref{spconstr}). To this purpose,
we introduce the intermediate
Dirac brackets (DB's) $\left\{\enskip,\enskip\right\} ^{*}$
associated to these constraints.
After some calculations one finds for the intermediate DB's among the
canonical variables the following expressions:
\begin{eqnarray}
\left\{ A_i^a(t,{\bf x}),\pi ^{j,b}(t,{\bf y})\right\} ^{*} &=&\frac 12%
\delta ^{ab}\delta _i^j\delta ({\bf x}-{\bf y})  \label{bone} \\
\left\{ A_i^a(t,{\bf x}),A_j^b(t,{\bf y})\right\} ^{*} &=&\frac{4\pi }s%
\delta ^{ab}\epsilon _{ij}\delta ({\bf x}-{\bf y})  \label{btwo} \\
\left\{ \pi ^{i,a}(t,{\bf x}),\pi ^{j,b}(t,{\bf y})\right\} ^{*} &=&\frac s{%
16\pi }\delta ^{ab}\epsilon ^{ij}\delta ({\bf x}-{\bf y})  \label{bthree}
\end{eqnarray}
Exploiting the DB's (\ref{bone})--(\ref{bthree}), we
obtain the relations:
\begin{eqnarray}
\left\{ \text{{\prova G}}^a(t,{\bf x}),A_i^b(t,{\bf y})\right\} ^*
&=&-D_i^{ab}(x)\delta ({\bf x}-{\bf y})\label{ffcone} \\
\left\{ \text{{\prova G}}[\psi ],A_i^a(t,{\bf x})\right\} ^*
&=& D_i^{ab}(x)\psi ^b({\bf x})\label{ffctwo} \\
\left\{ \text{{\prova G}}^a(t,{\bf x}),\text{{\prova G}}^b(t,{\bf y}%
)\right\}^*  &=&-f^{abc}\text{{\prova G}}^c(t,{\bf x})\delta ({\bf x-y)}
\label{ffcthree}
\end{eqnarray}
where $\text{\prova G}\left[ \psi \right] =\int d^2{\bf x}\text{\prova G}^a
(t,{\bf
x})\psi ^a({\bf x})$.
This shows that the $\text{\prova G}^a(t,{\bf x})$ are the generators of the $%
SU(N)$ gauge transformations as desired.

\subsection{Imposing the Coulomb gauge}
At this point, we are left with the constraints given by eq.
(\ref{fpconstr}) and by the Gauss law
(\ref{glaw}). 
The former constraint, which is first class and
involves the conjugate momentum of
$A_0^a$, can be ignored.
As a matter of fact, the field
$A_0^a$ just plays
the role of the Lagrange multiplier associated to the Gauss law
in the Hamiltonian (\ref{wrrr}) and has no dynamics.
From eqs. (\ref{ffcone})--(\ref{ffcthree}) it turns out that the Gauss law
(\ref{glaw}) is
a first class constraint. To make it second class,
we introduce the Coulomb gauge
fixing:
\begin{equation}
\partial _iA^{i,a}\approx 0  \label{coulombgauge}
\end{equation}
and the new extended Hamiltonian:
\begin{equation}
\check H_{CS}=\int d^2{\bf x}\left[ -A_0^a\text{{\prova G}}^a+\frac s{8\pi }%
A_i^a\partial ^iB^a+\lambda _0^a\pi ^{0,a}\right] \label{neham}
\end{equation}
From the condition  $\{\partial_i A^{i,a},\check H_{CS}\}^*
\approx 0$, we obtain an equation for $A_0^a$:
\begin{equation}
\partial^iD_i^{ab}A_0^b\approx 0\label{sfour}
\end{equation}
Moreover, the requirement
$\left\{ \partial^iD_i^{ab}A_0^b(x),\check H_{CS}
\right\}^*\approx 0$ determines the
Lagrange multiplier $\lambda_0$:
\begin{equation}
-\bigtriangleup\lambda_0^a
-\left\{\partial_i({\bf A}_i\times{\bf A}_0)^a,\check H_{CS}\right\}^*\approx0
\label{lzdet}
\end{equation}
In the above equation the symbol $\bigtriangleup$ denotes the two  dimensional
Laplacian $\bigtriangleup=-\partial_i\partial^i$ and
$$({\bf A}_i\times{\bf A}_0)^a\equiv f^{abc}A_i^bA_0^c$$
Another independent equation, which fixes the Lagrange multipliers $B^a$, is
provided by the requirement \.{\prova G}$^a\approx 0$:
\begin{equation}
\left\{{\text{\prova G}}^a,\check H_{CS}\right\}^*
\approx-\frac{s}{8\pi}D_i^{ab}\partial^i
B^b\approx 0\label{fixb}
\end{equation}
Let us notice that the above relations (\ref{glaw}), (\ref{coulombgauge}) and
(\ref{sfour})--(\ref{fixb}) are compatible with the equations of motion of the
gauge potentials:
\begin{equation} \epsilon^{ij}(D_i^{ab}A_j^{b}-\partial_jA_i^a)=0\label{eqone}
\end{equation}
\begin{equation} D_j^{ab}A_0^b-\partial_0A_j^a=0\label{eqtwo}
\end{equation}
As a matter of fact (\ref{eqone}) is equivalent to the condition
${\text{\prova G}}^a=0$. Moreover, multiplying for instance eq. (\ref{eqtwo})
with the differential operator $\epsilon_{ki}\partial^k$, we obtain the
relation:
$$\partial_0\partial^kA_k^a-\partial^kD_k^{ab}A_0^b=0$$
which is consistent with the Coulomb gauge
and the condition (\ref{sfour}) on $A_0^a$.

\subsection{The final Dirac brackets and their properties}
It is now possible to realize that the Gauss law
(\ref{glaw}) and the Coulomb gauge fixing
(\ref{coulombgauge}) form a set of second class constraints, so that we can
impose them
in the strong sense computing the final Dirac brackets $\left\{\enskip,
\enskip\right\}_{DB}$.
Putting
\[
\chi_1^a=\text{\prova G}^a\qquad\qquad\qquad
\chi_2^a=
\partial_i A^{i,a}
\]
with $\alpha,\beta=1,2$, and skipping all the technical details
of the calculations that can be found in ref.~\cite{frig},
we have:
\[
\left\{A_i^a({\bf x}),A_j^b({\bf y})\right\}_{DB}=
-\frac{4\pi}{s}\delta^{ab}\epsilon_{ij}\delta({\bf x}-{\bf y})+
\]
\begin{equation}
\frac{4\pi}{s}\epsilon_{ik}\partial_{\bf x}^kD_j^{bc}({\bf y})
\text{\prova D}^{ac}
({\bf x},{\bf y})-\frac{4\pi}{s}
\epsilon_{kj}D_i^{ac}({\bf x})\partial_{\bf y}^k\text{\prova D}^{cb}
({\bf x},{\bf y})\label{maincomrel}
\end{equation}
where
\begin{equation}
D^{ac}_i({\bf x})\partial_{\bf x}^i\text{\prova D}^{cb}({\bf x},{\bf y})=
\delta^{ab}\delta({\bf x}-{\bf y})\label{dstorta}
\end{equation}
After imposing the constraints (\ref{glaw}) and (\ref{coulombgauge}) in the
strong sense, the Hamiltonian $\check H_{CS}$ vanishes, but the commutation
relations (CR's) between the fields remain complicated.

Let us study the main properties of the above DB's.
\begin{itemize}
\item {\bf Antisymmetry}.
The antisymmetry of the right hand side of eq. (\ref{maincomrel}) is not
explicit, but can be verified with the help of the relation:
\begin{equation}
\label{propsym}
\text{\prova D}^{ab}({\bf x},{\bf y})=\text{\prova D}^{ba}({\bf y},{\bf x})
\end{equation}
The above identity is due to the fact that $\text{\prova D}^{ab}({\bf x},{\bf
y})$ is the Green function of the self-adjoint differential operator
defined in eq. (\ref{dstorta})~\cite{schwinger}.
Exploiting the above relation one finds that
\begin{equation}
\{ A_i^a({\bf x}), A_j^b({\bf y})\}_{DB}=-
\{ A_j^b({\bf y}), A_i^a({\bf x})\}_{DB}\label{antisym}
\end{equation}
as expected

\item {\bf Consistency with the Coulomb gauge constraint}.
The CR's (\ref{maincomrel}) are consistent
with the Coulomb gauge, i. e.:
$$\{ A_i^a({\bf x}), \partial^jA_j^b({\bf y})\}_{DB}=
\{ \partial^iA_i^a({\bf x}), A_j^b({\bf y})\}_{DB}=0$$

\item {\bf Covariance under the Poincar\'e group of transformations}.
The proof that the C--S theory in the Coulomb gauge is invariant under
the Poincar\'e group is not trivial
due to the complicated CR's (\ref{maincomrel}).
A  good strategy consists in evaluating the CR's among the generators
of the Poincar\'e group using the intermediate DB's
(\ref{bone})--(\ref{bthree}). In this way one finds that the Poincar\'e
algebra is not closed due to ``extra'' terms which are proportional to the
constraints.
For instance, for
the generators of the time and the space translations
we obtain the following result:
$$\{P_0, P_k\}^* = \int d^2x A_0^a \partial_k \text{\prova G}^a$$
where $\text{\prova G}$ is given in (\ref{glaw}).
Clearly, all these unwanted terms disappear after imposing the final
DB's (\ref{maincomrel}) and the CR's between the generetors of the
Poincar\'e group can be recovered.

\item{\bf Interactions}. For simplicity we have considered here
pure Chern--Simons field theories. However, we stress that the form of
the CR's (\ref{maincomrel}) remains unchanged also adding to the
lagrangian (\ref{lagrangian}) interactions of the kind
$L_I =\int d^3xA_\mu J^{\mu, a}$, where $J^{\mu, a}$ is a current
associated to matter fields. The only differences occur
in equations (\ref{glaw}), (\ref{lagdet}) and
(\ref{sfour})--(\ref{fixb}),
in which $J^{\mu, a}$ will appear as an external source.
For instance the Gauss law (\ref{glaw}) is modified as follows:
\begin{equation}
D_i^{ab}\pi ^{i,b}+\partial _i\pi
^{i,b}+J_0^a\approx 0\label{modgausslaw}
\end{equation}
\end{itemize}

\subsection{The abelian case}
The case of a Chern--Simons field theory with abelian gauge group $U(1)$ is
particularly instructive in order to understand the meaning of the CR's
(\ref{maincomrel}).
Indeed, in this case the Green function
$\text{\prova D}({\bf x},{\bf y})$ has the following simple expression:
\begin{equation}
\text{\prova D}({\bf x},{\bf y})=-{1\over 2\pi} {\rm log}|
{\bf x}-{\bf y}|\label{dsabelian}
\end{equation}
Let $U_\mu$ denote the abelian gauge fields.
Substituting the right hand side of equation (\ref{dsabelian}) in
(\ref{maincomrel}) and replacing the DB's with quantum commutators, we obtain:
$$[U_i(t,{\bf x}),U_j(t,{\bf y})]=0\label{abtriviality}$$
As a consequence the fields $U_\mu$ do not propagate.
This result is in agreement with the fact that the theory is topological
so that the fields have no dynamics. Indeed,
exploiting the Gauss law, the Coulomb gauge fixing and eqs.
(\ref{lagdet}), (\ref{sfour})--(\ref{fixb}), it is easy
to see that the the only possible solution of the equations of motion
is $U_i=U_0=\lambda_\mu=B=0$. 
On the other side, the triviality of the CR's holds also
in the presence of interactions, i. e. when the theory becomes no
longer topological and the solutions of the equations of motion
$U_\mu, \lambda_\mu, B$ are in general different from zero.

\subsection{The non-abelian case}
In non-abelian C--S field theories the equations of motion of the
constraints are nonlinear and can be solved only using a perturbative
approach. At the zeroth order, the Green function $\text{\prova
D}({\bf x}, {\bf y})$ is given again by eq. (\ref{dsabelian}).
Thus the CR's (\ref{maincomrel}) are zero at this order.
At higher orders, however, the right hand side of
eq. (\ref{maincomrel})
is in general different from zero, apart from the case in which there
are
no interactions. The vanishing of the CR's for the pure C--S theories,
proven at any perturbative order in the coupling constant $\frac 1s$
in~\cite{frig}, is in agreement with the fact that these theories
are topological.
Finally, we notice that the CR's (\ref{maincomrel}) are particularly
complicated with respect to the usual Yang--Mills field theories.
This is probably related to the fact that field theories coupled to
a C--S term exhibit a non-abelian statistics.

\subsection{Final Remarks}
\begin{itemize}
\item The C--S theories become
in the Coulomb gauge two dimensional
models. Only the fields $A_i^a$, for $i=1,2$, have in fact a dynamics, which
is governed by the commutation relations (\ref{maincomrel}).
Moreover, the latter do not contain time derivatives, so that the time
can be considered as an external parameter.

\item If no interactions with matter fields are present,
the CR's (\ref{maincomrel}) vanish at any perturbative order.
Thus the C--S field theories in the Coulomb gauge are not only finite,
but also free.
A natural question that arises at this point is if analogous conclusions
can be drawn for the covariant gauges.
For this reason it would be
interesting to repeat the procedure of canonical quantization developed
here 
also in this case.

\item 
If the interactions with other fields are switched on, the CR's
(\ref{maincomrel}) still remain trivial in the abelian case.
Thus, if we quantise the C--S theory replacing the DB's with quantum
commutators, we obtain that
\begin{equation}
\left[A_i^a({\bf x}),A_j^b({\bf y})\right]=0\label{quantumcr}
\end{equation}
In the non-abelian case the above relation is valid
only at the zeroth order
in the coupling constant $\frac 1s$, while
at higher orders the CR's do not vanish and are rather complicated.
Let us notice that there is no contradiction between eq.
(\ref {quantumcr}) and the fact that, starting from the Lagrangian
(\ref{lagrangian}),
it is possible to derive a non-zero propagator for  the
C--S fields.
In  fact, going from the Hamiltonian normal-ordered formalism to the
Lagrangian time-ordered formalism it is know that contact
terms may arise, which contain distributions in the time variable.
Indeed, the components of the propagator computed in the next section will
have exactly the form of contact terms of this kind~\cite{flprep}.

\item
Finally, we have shown that the CR's (\ref{maincomrel})
are perfectly well defined and do not lead
to ambiguities in the quantization of the
C--S models in the Coulomb gauge. They are
consistent with the constraints and the Poincar\'e covariance of
the theory. Moreover, in the pure C--S field theory the CR's
vanish at any perturbative order in $\frac 1s$.

\end{itemize}


\section{Perturbative Analysis of the C--S Field Theory in the Coulomb
Gauge}

\subsection{Derivation of the Feynman rules}
In this section we consider the following gauge fixed
C--S action:
\begin{equation}
S_{CS}=S_0+S_{GF}+S_{FP}
\label{action}
\end{equation}
whith
\begin{equation}
S_0=\frac s{4\pi }\int
d^3x\epsilon ^{\mu \nu \rho }\left( \frac
12A_\mu ^a\partial _\nu A_\rho ^a-\frac 16f^{abc}A_\mu ^aA_\nu ^bA_\rho
^c\right)  \label{csaction}
\end{equation}

\begin{equation}
S_{GF}=\frac {is}{8\pi \lambda }\int d^3x\left( \partial
_iA^{a\,i}\right) ^2  \label{gf}
\end{equation}
and 
\begin{equation}
S_{FP}=i\int
d^3x\,\overline{c}^a\partial _i\left( D^i\left[
A\right] c\right) ^a  \label{fp}
\end{equation}
In (\ref{gf}) $\lambda $ is a real gauge fixing parameter.
With respect to the previous section the metric is now Euclidean and
is of the form $g_{\mu\nu}=\mbox{\rm  diag}(1,1,1)$.
Moreover, the covariant derivatives are now defined as follows
$$D_\mu^{ab} \left[ A\right] =\partial _\mu \delta ^{ab}-f^{abc}A_\mu
^c$$
In eq. (\ref{action}) the Coulomb gauge constraint is weakly imposed
and the proper Coulomb gauge fixing
\begin{equation}
\partial _iA^{a\,i}=0  \label{gaugefix}\qquad\qquad\qquad i=1,2
\end{equation}
of the previous section
is recovered setting $\lambda=0$ in eq. (\ref{gf}).


From (\ref{action}) the components of the gauge field
propagator
$G_{\mu \nu }^{ab}(p)$ in the Fourier space
are given by:
\begin{equation}
G_{jl}^{ab}(p)=
-\delta ^{ab}\frac{4\pi \lambda }s\frac{p_ip_l}{{\mbox{\rm\bf p}}^4}
\label{gjl}
\end{equation}

\begin{equation}
G_{j0}^{ab}(p)=
\delta ^{ab}
\left(
\frac{4\pi }s\epsilon _{0jk}\frac{p^k}{\mbox{\rm\bf p}^2}-
\frac{4\pi \lambda }s
\frac{p_jp_0}{\mbox{\rm\bf p}^4}
\right)
\label{gjo}
\end{equation}

\begin{equation}
G_{0j}^{ab}(p)=
-\delta ^{ab}
\left( 
\frac{4\pi }s
\epsilon _{0jk}
\frac{p^k}{\mbox{\rm\bf p}^2}+
\frac{4\pi \lambda }s\frac{p_0p_j}{\mbox{\rm\bf p}^4}\right) 
\label{goj}
\end{equation}

\begin{equation}
G_{00}^{ab}(p)=
-\delta^{ab}
\frac{4\pi \lambda }s
\frac{p_0^2}{\mbox{\rm\bf p}^4}
\label{goo}
\end{equation}
where $\mbox{\rm\bf p}^2=p_1^2+p_2^2$.
Let us notice that the variable $p_0$ appears only in the longitudinal
contributions to the propagator and disappears after choosing the
proper
Coulomb gauge.
Also the ghost propagator $G_{gh}^{ab}(p)$ is independent on $p_0$:
\begin{equation}
G_{gh}^{ab}(p)=\frac{\delta ^{ab}}{\mbox{\rm\bf p}^2}  \label{ggh}
\end{equation}
Finally, the three-gluon vertex and the ghost-gluon vertex
are respectively given by:
\begin{equation}
V_{\mu _1\mu _2\mu _3}^{a_1a_2a_3}(p,q,r)=-\frac{is}{3!4\pi }(2\pi
)^3f^{a_1a_2a_3}\epsilon ^{\mu _1\mu _2\mu _3}\delta ^{(3)}(p+q+r)
\label{aaa}
\end{equation}
and
\begin{equation}
V_{\mathrm{gh\thinspace }i _1}^{a_1a_2a_3}(p,q,r)=-i(2\pi )^3\left(
q\right) _{i_1}f^{a_1a_2a_3}\delta ^{(3)}(p+q+r)  \label{acc}
\end{equation}
In the above equation we have only given the spatial components of the
ghost-gluon vertex.
From eq. (\ref{fp}), it is in fact easy to realize that
in the Coulomb gauge
its temporal component is zero.
As we see, the presence of $p_0$ remains confined in the vertices
(\ref{aaa})--(\ref{acc})
and it is trivial because it is concentrated in the Dirac $\delta $%
--functions expressing the momentum conservations. As a consequence,
the CS field theory can be considered as a two dimensional model in
the
proper Coulomb gauge.

\subsection{Potential divergences}
At this point, we study the divergences that may arise in the
computation of the Feynman diagrams.
The potential divergences are
of three kinds: ultraviolet, infrared and spurious.

\begin{enumerate}
\item  Ultraviolet divergences (UV). The naive power counting gives the
following degree of divergence $\omega (G)$ for a given Feynman diagram $G$: 
\begin{equation}
\omega (G)=3-\delta -E_B-\frac{E_G}2  \label{napoco}
\end{equation}
with \footnotemark{}\footnotetext{We use here the same notations of
ref.~\cite{itzu}}

\begin{enumerate}
\item  $\delta =$ number of momenta which are not integrated inside the loops

\item  $E_B=$ number of external gluonic legs

\item  $E_G=$ number of external ghost legs
\end{enumerate}

Eq. (\ref{napoco}) shows that UV divergences are possible in the
two and three point functions, both with gluonic or ghost
legs. Moreover, there is also a possible logarithmic divergence in the case
of the four point interaction among two gluons and two ghosts.
In principle, we had to introduce a regularization for these divergences
but in practical calculations this is not necessary.
As a matter of fact, we will see below that there are no
UV divergences in the quantum corrections of the Green functions.

\item  Infrared (IR) divergences. 
The pure C--S field theories are known to be free of infrared
divergences~\cite{hageno} so that there is no need to discuss them.

\item  Spurious divergences.
These singularities appear because the propagators
(\ref{gjl})--(\ref{ggh}) are undamped in the time direction and are
typical of the Coulomb gauge.
To regularize spurious divergences
of this kind, 
it is
sufficient to introduce a cutoff $\Lambda _0>0$ in the domain of integration
over the variable $p_0$: 
\begin{equation}
\int_{-\infty }^\infty dp_0\rightarrow \int_{-\Lambda _0}^{\Lambda _0}dp_0
\label{spureg}
\end{equation}
The physical situation is recovered in the limit $\Lambda _0\rightarrow
\infty $. As we will see, this regulatization does not cause
ambiguities in the evaluation of the
radiative corrections at any loop order.
In fact, the integrations over the temporal components of the
momenta inside the loops turn out to be trivial and do not interfere
with
the integrations over the spatial components.
\end{enumerate}

\noindent

\subsection{Perturbative analysis at one loop order}

In this Section we compute the $n-$point
correlation functions of C--S field theories at one loop order.
From now on, we choose for simplicity
the proper Coulomb gauge, setting
$\lambda=0$ in eq. (\ref{gf}).
In this
gauge the gluon-gluon propagator has only two nonvanishing components: 
\begin{equation}
G_{j0}(p)=
-G_{0j}(p)=
\delta^{ab}
\frac{4\pi }s
\epsilon _{0jk}\frac{p^k}{
\mbox{\rm\bf p}^2}  \label{gjopcg}
\end{equation}

The following observation greatly reduces the number of diagrams to be
evaluated:

\begin{description}
\item[Observation:] 
Let $G^{(1)}$ be a one particle irreducible (1PI) Feynman diagram containing
only one closed loop. Then all
internal lines of $G^{(1)}$ are either ghost or gluonic lines.
\end{description}

A proof of the above observation can be found in ref.~\cite{flpert}.
An important consequence is that,
at one loop, the only non--vanishing diagrams occur when all the external
legs are gluonic. Hence we have to evaluate only the diagrams describing the
scattering among $n$ gluons.

This can be done as follows. First of all, we
consider the  diagrams
with internal gluonic lines.
After suitable
redefinitions of the indices and of the momenta, it is possible to see that
their total contribution is given by:
\begin{eqnarray}
V_{i_1...i_n}^{a_1...a_n}
\left(
1;p_1,...,p_n
\right)=
C\left[ -i
\left(2\pi
\right)^3
\right]^n
\frac{n!(n-1)!}2\delta^{(2)}
(\mbox{\rm\bf p}_1+...+
\mbox{\rm\bf p}_n) &&  \label{vone} \\
f^{a_1b_1^{\prime }c_1^{\prime }}f^{a_2b_2^{\prime }b_1^{\prime
}}...f^{a_nc_1^{\prime }b_{n-1}^{\prime }}\int d^2\mbox{\rm\bf q}_1
\frac{\left[
q_1^{i_1}...q_n^{i_n}+q_1^{i_2}\ldots q_j^{i_{j+1}}
\ldots q_{n-1}^{i_n}q_n^{i_1}\right] }{
\mbox{\rm\bf q}%
_1^2...\mbox{\rm\bf q}_n^2} &&  \nonumber
\end{eqnarray}
where $C=\left( 2\Lambda _0\right) ^{2n}$ is a finite constant coming from
the integration over the zeroth components of the momenta and
\begin{equation}
\begin{array}{cccc}
q_2= & q_1+p_1+p_n+p_{n-1}+ & \ldots & +p_3 \\ 
\vdots & \vdots&\ddots&\vdots \\ 
q_j=&q_1+p_1+p_n+p_{n-1}+&\ldots&+p_{j+1} \\ 
\vdots & \vdots&& \\ 
q_n=&q_1+p_1&&
\end{array}
\label{qform}
\end{equation}
for $j=2,\ldots,n-1$.

The case of the Feynman diagrams containing ghost internal lines
is more complicated. After some work, it is possible to
distinguish two different contributions to the Green functions
with $n$ gluonic legs that cannot be reduced into one by
renaming indices and momentum variables:
\begin{eqnarray}
V_{i_1...i_n}^{a_1...a_n}\left( 2a;p_1,...,p_n\right)=-C\left[ -i\left(
2\pi \right) ^3\right]^n\frac{n!(n-1)!}2
&&  	\nonumber \\
\delta^{(2)}(\mbox{\rm\bf p}_1+...%
\mbox{\rm\bf p}_n)
f^{a_1b_1^{\prime }c_1^{\prime }}f^{a_2b_2^{\prime }b_1^{\prime
}}...f^{a_nc_1^{\prime }b_{n-1}^{\prime }}\int d^2\mbox{\rm\bf q}_1\frac{%
q_1^{i_1}...q_n^{i_n}}{\mbox{\rm\bf q}_1^2...
\mbox{\rm\bf q}_n^2}\label{vtwoa}&&
\end{eqnarray}
and 

\[
V_{i_1...i_n}^{a_1...a_n}\left( 2b;p_1,...,p_n\right) = 
-C[-i(2\pi)^3]^n
\frac{n!(n-1)!}2f^{a_1b_1^{\prime
}c_1^{\prime }}f^{a_2b_2^{\prime }b_1^{\prime }}...f^{a_nc_1^{\prime
}b_{n-1}^{\prime }}
\]
\begin{equation}
\delta ^{(2)}(\mbox{\rm\bf p}_1+...+\mbox{\rm\bf p}_n) 
\int d^2\mbox{\rm\bf q}_1
\frac{q_n^{i_1}q_1^{i_2}\ldots q_j^{i_{j+1}}\ldots q_{n-1}^{i_n}}
{(\mbox{\rm\bf q}_1')^2...
(\mbox{\rm\bf q}_n')^2}\label{finvtb}
\end{equation}
In the above equations the variables $q_2,\ldots,q_n$ and the constant
$C$ are the same as in eq.
(\ref{vone}).

As it is possible to see from eqs. (\ref{vone}).(\ref{vtwoa}) and
(\ref{finvtb}),
the only nonvanishing components of the $n$ points functions
are those for which all tensor indices
$\mu_1,\ldots,\mu_n$
are
spatial.
We notice here that eq. (\ref{finvtb}) has been obtained after a shift
of the integration variable $q_1$.
However,  it is not difficult to verify that
the right hand sides of eqs. (\ref{vone})--(\ref{finvtb})
are neither IR nor UV divergencent for $n\ge 3$, so that a shift of
$q_1$ is not dangerous.
At this point we can sum 
equations (\ref{vone}).(\ref{vtwoa}) and
(\ref{finvtb}) together.
It is easy to realize that the total result is zero, i. e.:

\begin{equation}
V_{i_1...i_n}^{a_1...a_n}\left( 1;p_1,...,p_n\right)
+V_{i_1...i_n}^{a_1...a_n}\left( 2a;p_1,...,p_n\right)
+V_{i_1...i_n}^{a_1...a_n}\left( 2b;p_1,...,p_n\right) =0 \label{finres}
\end{equation}

For $n \ge 3$ this result is regularization independent since the
Feynman integrals are IR and UV convergent.

Only the case $n=2$ needs some more care and will be treated separately.
After a few calculations one obtains that
for $n=2$ the total contribution to the gluonic propagator to one loop
is given by:
%
\[
V_{ij}^{ab}\left( 1;p_1,p_2\right) +V_{ij}^{ab}\left(
2a;p_1,p_2\right) +V_{ij}^{ab}\left( 2b;p_1,p_2\right) = 
\]
\begin{equation}
\left( 2\pi \right) ^6\left( 2\Lambda _0\right) ^2N\delta ^{ab}\delta
^{(2)}(\mbox{\rm\bf p}_1+\mbox{\rm\bf p}_2)
\int d^2\mbox{\rm\bf q}\frac{\left[ q_{i}
(p_1)_{j}-q_{j}(p_1) _{i}
\right] }
{\mbox{\rm\bf q}^2\left( \mbox{\rm\bf q}+\mbox{\rm\bf p}_1\right) ^2}
\label{cru}
\end{equation}
The integrand appearing
in the rhs of (\ref{cru})
is both IR and UV
finite.  Moreover, 
a simple computation shows that 
the integral over $\mbox{\rm\bf q}$ in (\ref{cru})
is zero.
As a consequence, there are no
contributions to the Green functions at one loop.

\subsection{The higher order radiative corrections}
\begin{figure}
\includegraphics{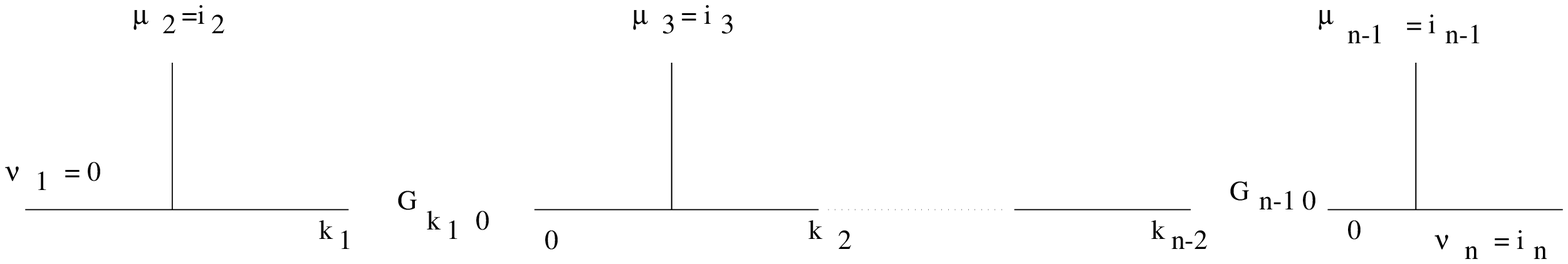}
\vspace{0.25in}
\caption{This  figure shows that in an arbitrary tree diagram
$T_{\nu_1\nu_2\ldots\nu_{n-1}\nu_n}$
constructed in terms of the
gauge fields propagator (\ref{gjopcg})  and the
three gluon vertex (\ref{aaa}), only one component in the
space-time indices $\nu_i$,
$i=1,\ldots,n$, can be temporal.}
\label{figtr}
\end{figure}

Now we are ready to consider the higher order corrections of the $n$
points Green functions.
At two loop, a general Feynman diagram $G^{(2)}$ can be obtained
contracting two legs of a tree diagram $G^{(0)}$ with
two legs of a one loop diagram $G^{(1)}$.
As previously seen, the latter have only gluonic
legs and their tensorial indices are all spatial.
Consequently, in order to perform the contractions by means of the propagator
(\ref{gjopcg}), there should exist one component of 
$G^{(0)}$ with at least two temporal indices,
but this is impossible. To convince
oneself of this fact, it
is sufficient to look at fig. (\ref{figtr}) and related comments.
The situation does not improve
if we build $G^{(0)}$  exploiting also the ghost-gluon vertex
(\ref{acc}), because it has no temporal component.
As a consequence, all the Feynman graphs vanish identically at two loop order.
Let us notice that it is possible to verify their vanishing
directly, since
the number of two loop diagrams
is relatively small in the Coulomb gauge and one has just to contract the
space-time indices without performing the integrations over the internal
momenta.
However, this
procedure is rather long and will not be reported here.

The proof that also higher order diagrams vanish can be done by
induction. First of all,
a diagram with $N+1$ loops $G^{(N+1)}$
has at least
one subdiagram $G^{(N)}$
containing $N-$loops.
Supposing that $G^{(N)}$ is identically equal to
zero because it cannot be constructed with the
Feynman rules (\ref{ggh})--(\ref{acc}) and (\ref{gjopcg}), also $G^{(N+1)}$
must be zero.
As we have seen above, there are no Feynman diagrams for $N=2$.
This is enough to prove by induction that
C--S field theories have  no radiative corrections
in the Coulomb gauge
for any value of $N$.

\subsection{Final remarks}
\begin{itemize}

\item
In the Coulomb gauge the C--S field theories do not have quantum
corrections
at any loop order, has as been shown by explicit computations.

\item
IR and UV divergences are absent in the calculations of the
Feynman diagrams.
Only spurious singularities are present,
related to the fact
that the propagators are undamped in the time direction.
They are similar to the singularities observed in the four dimensional
Yang--Mills field theories~\cite{taylor},
but in the C--S case appear in a milder form.
In fact, after introducing
the regularizarion (\ref{spureg}) and integrating over the time
component of the momenta in a given amplitude, the total
contribution at any loop order reduces to
an overall constant factor. The remaining calculations
consist of finite two dimensional Feynman
integrals over the space variables.
As a consequence, the results obtained here are regularization
independent.

\item
The vanishing of the quantum contributions described in Section 3
is a peculiarity of the Coulomb gauge that does not strictly depend the
fact that the C--S field theories are topological. In fact, finite
remormalizations of the fields and of the coupling constant $s$ are
always possible as it happens in the case of the covariant gauges.
An analogous situation in which there are no radiative corrections
occurs in the light cone gauge in the presence of a
boundary. In that case, radiative corrections arise in principle
due to the interactions
of the fields with the boundary~\cite{empi} but all the related
Feynman diagrams
vanish identically.

\item
The C--S field theories in the proper Coulomb
gauge can be considered as two dimensional models.
This has been shown in the
previous section and
has been confirmed here by the fact that the
dependence on the the time component of the momenta in the propagators
and vertices is trivial.

\item
Contrary to what happens using the covariant gauges or the axial
gauges, the Coulomb gauge can easily be applied also when
space-times with non-trivial spatial section are considered,
like for instance a Riemann surface.
The absence of radiative corrections 
particularly useful in this
case, where the momentum representation does not exist and this the
evaluation of Feynman diagrams becomes forbiddenly difficult.

\item
The C--S field
theories
can be considered
as a good laboratory in order to study the possible remedies
of pathologies that appear in similar ways in the more complicated
four dimensional gauge field theories.
For example, it would be interesting
to apply to the Yang--Mills case
the regularization (\ref{spureg}) introduced here for the spurious
singularities.
Let us
notice that a different regularization
has been recently proposed in~\cite{leiwil}.
\end{itemize}

\section{Chern--Simons field theories in the Coulomb Gauge on Curved
Space--Times}

In this section we
consider a manifold $M_3$ with a Robertson-Walker metric and
Euclidean signature
of the kind
\begin{equation}
g_{00}=1\qquad\qquad g_{z\bar z}(z,\bar z,t)=g_{z\bar z}(z,\bar z,t)=a(t)
h(z,\bar z)\qquad\qquad g^{z\bar z}g_{z\bar z}=1
\label{compmetric}
\end{equation}
$g^{z\bar z}$ is the metric on a Riemann surface
$\Sigma_g$ of genus $g$ and
$z$ and $\bar z$ are local coordinates on
$\Sigma_g$:
We suppose that $a(t)>0$ for each values of the time $t$.
Thus $M_3$  correspond to an expanding universe  having the
Riemann surface $\Sigma_g$ as spatial section.

The gauge fixed
Chern-Simons action (\ref{action}) becomes in complex coordinates
$S_{\rm CS}=S_{\rm free}+S_{\rm int}$, where:
$$S_{\rm free}=
\int_{M_3}
d^2zdt
\left[
2i
\left(
A_0^a\partial_{\bar z}A_z^a+
A_z^a\partial_0A_{\bar z}^a+
A_{\bar z}^a\partial_zA_0^a-
{\rm c.c.}
\right)
\right.$$
\begin{equation}
\left.
+\left(
a(t)\lambda
\right)^{-1}
g^{\bar z z}
(\partial_zA_{\bar z}^a+
\partial_{\bar z}A_z^a)^2
+2\bar c^a\partial_z\partial_{\bar z}c^a\right]\label{csfree}
\end{equation}
\begin{equation}
S_{\rm
int}=\int_{M_3}d^2zdt\left[\epsilon^{\mu\nu\rho}f^{abc}
A_\mu^aA_\nu^bA_\rho^c -f^{abc}\bar c^a\left(A_z^b\partial_{\bar
z}+A_{\bar z}^b\partial_z\right)c^c\right]
\label{csint}
\end{equation}
and $d^2z={1\over 2i}dz\wedge d\bar z$.
The factor $2i$ in eq. (\ref{csfree}) comes from the form of the
$\epsilon^{\mu\nu\rho}$ tensor in complex coordinates.
In fact, the Levi-Civita tensor $[\epsilon]^{\mu\nu\rho}=g^{-{1\over 2}}
\epsilon^{\mu\nu\rho}$ becomes in these coordinates:
\begin{equation}
[\epsilon]^{0z\bar z}=-2ig^{z\bar z} a^{-1}(t)
\label{levicivita}
\end{equation}
All the other components can be obtained from eq. (\ref{levicivita})
permuting the indices $0$, $z$ and $\bar z$ and changing the sign
according to the order of the permutation.
In this Section it will be useful to denote a sum over the complex
indices with the first letters
of the Greek alphabet $\alpha,\beta,\gamma$ and so on.
For example, the Coulomb gauge condition becomes now $\partial^\alpha
A_\alpha^a=0$. Using the metric (\ref{compmetric})
to rise and lower the indices,
this equation reads:
\begin{equation}
\partial_zA_{\bar z}^a+\partial_{\bar z}A_z^a=0\label{complgauge}
\end{equation}
Eq. (\ref{complgauge})
 does not contain the metric explicitely. This means that the
Coulomb gauge condition is compatible with the transition functions at
the intersections of the open sets covering
the Riemann surface $\Sigma_g$.
Therefore eq. (\ref{complgauge}) is globally valid on
$M_3$.\smallskip
The gauge fields $(A_z^a,A_{\bar z}^a,A_0^a)$ are connections on the
trivial principal bundle $$P(M_3,{\rm SU(N)})=M_3\otimes{\rm SU(N)}$$
This bundle is trivial due to the fact that SU(N) is a simply connected
Lie group.
One can show as in the flat case that the Coulomb gauge (\ref{complgauge}) is
a good gauge fixing without Gribov ambiguities \cite{gribov}
at least in the
perturbative approach (see ref. \cite{ffunp} for details.
We are now ready to compute the propagators of the gauge
fields
$$G^{ab}_{\mu\nu}(z,w;t,t')=<A^a_\mu(z,\bar z,t)A^b_\nu(w,\bar w,t)>$$
where now $\mu,\nu=0,z,\bar z$.
The equations satisfied by the above propagator are:
\begin{equation}
-4i\partial_z
G^{ab}_{\bar z0}(z,w;t,t')
+4_i\partial_{\bar
z}
G^{ab}_{z0}(z,w;t,t')
={8\pi\over s}\delta^{ab}\delta^{(2)}_{z\bar
z}(z,w)\delta(t-t')\label{propI}
\end{equation}
$$-4i\partial_{\bar
z}
G^{ab}_{0w}(z,w;t,t')
+4i\partial_0
G^{ab}_{\bar z w}(z,w;t,t')
-2{a^{-1}(t)\over \lambda}\partial_{\bar
z}\left[g^{z\bar z}\partial_{\bar z}
G^{ab}_{zw}(z,w;t,t')
+\right.$$
\begin{equation}
\left.+
g^{z\bar z}\partial_z
G^{ab}_{\bar z w}(z,w;t,t')
\right]={8\pi\over s}\delta^{ab}\delta^{(2)}_{\bar z w}
(z,w)\delta(t-t')\label{propII}
\end{equation}
Another equation can be obtained from (\ref{propII})
permuting the indices $z$ and $\bar z$ and substituting the index $w$
with $\bar w$.
There are still other relations relating the various components of the
propagators together:
\begin{equation}
\partial_z
G^{ab}_{\bar z \alpha}(z,w;t,t')
-
\partial_{\bar z}G^{ab}_{z\alpha}(z,w;t,t')
=0\label{propIII}
\end{equation}
$$-4i\partial_\alpha
G^{ab}_{00}(z,w;t,t')
+4i\partial_0
G^{ab}_{\alpha0}(z,w;t,t')
-$$
\begin{equation}
-{a^{-1}(t)\over
\lambda}\partial_\alpha\left[ g^{z\bar z}\partial_{\bar
z}
G^{ab}_{z0}(z,w;t,t')
+g^{z\bar z}\partial_z
G^{ab}_{\bar z 0}(z,w;t,t')
\right]=0\label{propIV}
\end{equation}
where $\alpha=w,\bar w$. Eq. (\ref{propIII}) implies that the propagators
$G_{z\bar z}^{ab}(z,w;t,t')$ and $G_{\bar z z}^{ab}(z,w;t,t')$
do not have transverse components..
Finally we have:
$$-4i\partial_{\bar z}
G^{ab}_{0\bar w}(z,w;t,t')
+4i\partial_0
G^{ab}_{\bar z\bar w}(z,w;t,t')
-$$
\begin{equation}
-{a^{-1}(t)\over \lambda}\partial_{\bar z}\left[g^{z\bar
z}\partial_{\bar z}
G^{ab}_{z\bar w}(z,w;t,t')
+g^{z\bar
z}\partial_z
G^{ab}_{\bar z\bar w}(z,w;t,t')
\right]=0\label{propV}
\end{equation}
Again it is possible to get another independent relation from eq.
(\ref{propV}) interchanging the two indices $z$ and $\bar z$ and substituting
$\bar w$ with $w$.
Eqs. (\ref{propI}--\ref{propV})
are the equivalent of the equations defining the
propagator in the
flat case. However, they are still uncomplete, because in deriving
them we have neglected the zero mode contributions. In fact,
we should remember that due to  a theorem stating that the
total charge on a Riemann surface (like in any other two dimensional
compact manifold) is always zero, an isolated $\delta$ function
$\delta^{(2)}(z,w)$ is not allowed. Therefore, in the right hand sides
of eqs. (\ref{propI}--\ref{propII}) there must be also terms
containing
zero modes, whose expressions
will be uniquely determined below.
Since it is very difficult to solve equations
(\ref{propI}--\ref{propV})
for any value of $\lambda$,
we choose here the
proper Coulomb gauge taking the limit
$\lambda\rightarrow 0$. In this case drastic simplifications occur,
so that the above equations reduce to the following
two relations:
\begin{equation}
\partial_z
G^{ab}_{\bar z0}(z,w;t,t')
-\partial_{\bar
z}
G^{ab}_{z0}(z,w;t,t')
={4\pi i\over
s}\delta^{ab}\delta^{(2)}(z-w)\delta(t-t')+{\rm zero}\enskip{\rm
modes}
\label{simpleone}
\end{equation}
\begin{equation}
\partial_{\bar
z}
G^{ab}_{z0}(z,w;t,t')
+\partial_z
G^{ab}_{\bar z 0}(z,w;t,t')
=0\label{simpletwo}
\end{equation}
These equations describe exactly the main requirement of the Coulomb
gauge, i.e. the fact that only the transverse fields in the two
dimensional spatial section $\Sigma_g$ of $M_3$ propagate.
The transverse fields in complex coordinates satisfy in fact
the following condition: $A_z^a=\overline{(A_z^a)}=-A_{\bar z}^a$.
The solution of eqs. (\ref{simpleone}) and (\ref{simpletwo}) is provided by the
following Green functions:
\begin{equation}
<A_z^a(z,t)A_0^b(w.t')>={2\pi i\over s}
\delta^{ab}\partial_zK(z,w)\delta(t-t')\label{azaz}
\end{equation}
and
\begin{equation}
<A_{\bar
z}^a(z,t)A_0^b(w,t')>=-{2\pi i\over s}
\delta^{ab}\partial_zK(z,w)\delta(t-t')
\label{azbaz}
\end{equation}
where $K(z,w)$ is the usual propagator of the scalar fields on a Riemann
surface satisfying the equations (see ref. \cite{vv}
for more details):
\begin{equation}
K(z,w)=\delta_{z\bar z}^{(2)}(z,w)+{g_{z\bar z}\over
\int_{\Sigma_g} d^2ug_{u\bar u}}\label{dzdzbk}
\end{equation}
\begin{equation}
\partial_z\partial_{\bar w}K(w,z)=-\delta^{(2)}_{z\bar w}(z,w)
+\bar\omega_i(\bar
z)\left[{\rm Im}\enskip \Omega\right]^{-1}_{ij}\omega_j(w)
\label{dzdwbk}
\end{equation}
\begin{equation}
\int_{\Sigma_g}d^2zg_{z\bar z}K(z,w)=0\label{ik}
\end{equation}
In eq. (\ref{dzdwbk}) the $\omega_i(z)dz$, $i=1,\ldots,g$, denote the usual
holomorphic differentials and $\Omega_{ij}$ represents the period
matrix.
It is important to stress here that $K(z,w)$ is a singlevalued function
on $\Sigma_g$.
Using the propagators (\ref{azaz}) and (\ref{azbaz}) it is easy to see that eq.
(\ref{simpletwo}) is trivially satisfied.
Therefore, the Coulomb gauge requirement (\ref{complgauge})
is fulfilled and the above defined propagators describe exactly the transverse
components of the gauge fields.
Still there is an ambiguity in the solutions (\ref{azaz}) and
(\ref{azbaz}) due to
the zero mode sector of the fields $A_z^a$ and $A_{\bar z}^a$.
In order to remove this ambiguity, we have to require that the above
propagators are singlevalued along the nontrivial homology cycles of the
Riemann surface. Otherwise, the propagators are not well defined on
$M_3$, but in one of its coverings.
Therefore, the propagators should obey the following relations:
\begin{equation}
\oint_{\gamma}dz<A_z^a(z,t)A_0^b(w,t')>=
\oint_{\bar \gamma}d\bar z<A_{\bar
z}^a(z,t)A_0^b(w,t')>=0\label{zeroholonomy}
\end{equation}
along any nontrivial homology cycles $\gamma$.
Due to the properties of singlevaluedness of the Green function $K(z,w)$,
eq. (\ref{zeroholonomy}) is trivially satisfied by the propagators given in
eqs. (\ref{azaz}) and
(\ref{azbaz}).
In this way these two propagators are well defined and also the freedom
in the zero mode sector is removed.
Now we insert their expressions in eq. (\ref{simpleone}) in order to get
the exact form of the zero mode terms appearing
in the right hand side of this
equation:
$$
\partial_z
G_{\bar z0}^{ab}(z,w;t,t')
-\partial_{\bar
z}
G_{z0}^{ab}(z,w;t,t')
=$$
\begin{equation}
{4\pi i\over
s}\delta^{ab}\delta^{(2)}(z,w)\delta(t-t')+ {4\pi is}{g_{z\bar z}
\over \int_{\Sigma_g} d^2ug_{u\bar u}}\delta(t-t')\label{simplezm}
\end{equation}
The fact that the propagators in the Coulomb gauge must obey eq.
(\ref{zeroholonomy}) can be understood also decomposing the fields by means of
the Hodge decomposition of the gauge fields in a coexact, exact and
harmonic part:
\begin{equation}
A_z^a=i\partial_z\varphi^a+\partial_z\rho^a+A_z^{\rm har}\label{azdec}
\end{equation}
\begin{equation}
A_{\bar z}^a=i\partial_{\bar z}\varphi^a+\partial_{\bar z}
\rho^a+A_{\bar z}^{\rm har}\label{azbdec}
\end{equation}
$\phi^a$ and $\rho^a$ represent two real scalar fields.
The above decomposition is allowed since the gauge invariance has been
completely fixed by the choice of the Coulomb gauge, at least in the
perturbative approach, and the $G-$bundle $P(M_3,{\rm SU(N)})$ is
trivial as we previously remarked.
In the Coulomb gauge, the only components of the gauge fields which are
allowed to propagate are the coexact differentials, i.e. the
$1-$forms obtained differentiating the scalar fields $\varphi^a$ in eqs.
(\ref{azdec}) and (\ref{azbdec}). Therefore, the requirement 
(\ref{zeroholonomy})
is a pure consequence of the fact that
the coexact forms have vanishing holonomies around the nontrivial
homology cycles.

Let us notice that the zero mode term appearing in the right hand side
of eq. (\ref{simplezm}) is totally irrelevant.
To eliminate it it is sufficient to introduce new gauge fields, let
say $\tilde A_z$, $\tilde A_{\bar z}$, differing from the old ones
by the fact that they are normalized to zero at a point $(0,0)$ of the
Riemann surface\footnotemark{}\footnotetext{On $M_3$ this implies that
the
new fields are
normalized to zero along the whole line of the time. This is possible to
do since the three dimensional manifold
is flat in the time direction.}:
\begin{equation}
\tilde A_z^a(z,\bar z,t)=A_z^a(z,\bar z,t)-A_z^a(0,0,t)\label{newaz}
\end{equation}
\begin{equation}
\tilde A_{\bar z}^a(z,\bar z,t)=
A_{\bar z}^a(z,\bar z,t)-A_{\bar z}^a(0,0,t)\label{newazb}
\end{equation}
Using the above new fields it is easy to check that the second term
in the right hand
side of eq. (\ref{simplezm}), which is a zero mode contribution, cancels out.

We finish this Section providing the explicit form of the other correlation
functions of Chern-Simons field theory.
The propagator of the ghost fields becomes:
$$G_{\rm gh}^{ab}(z,w;t,t')=\delta^{ab}K(z,w)\delta(t-t')$$
The vertex coming from the cubic interaction between the gauge fields
reads instead:
$$V_{z_100}^{abc}(z_1,z_2,z_3;t,t',t'')=
{2\pi^2 is\over
3}\int_{\Sigma_g}
d^2zf^{abc}\partial_{z_1}K(z_1,z)\left[\partial_zK(z_2,z)\partial_{\bar
z}K(z_3,z)-\right.$$
$$\left.\partial_{\bar
z}K(z_2,z)\partial_zK(z_3,z)\right]\delta(t-t'')\delta(t'-t'')$$
The simple integration in the variable $t$ has been already carried out
in the above expression of the vertex.
The component
$V^{abc}_{\bar z_1 00}(z_1,z_2,z_3;t,t',t'')$ of the vertex can be
simply obtained replacing the derivative $\partial_{z_1}$ in the above
equation with its complex conjugate.
Finally, the vertex describing the interaction between ghost and
gauge fields has
only one component which is given by:
$$V_{0\enskip {\rm gh}}^{abc}(z_1,z_2,z_3;t,t',t'')={2\pi
i\over sa(t)}
\int_{M_3}d^2zf^{abc}K(z_1,z)\left[\partial_zK(z_2,z)\partial_{\bar
z}K(z_3,z)\right.$$
$$\left.-\partial_{\bar z}K(z_2,z)\partial_zK(z_3,z)\right]
\delta(t-t')\delta(t'-t'')$$
It is easy to check that the above expressions of the
vertices are real as it should be.

\section{Conclusions}
In summary, our study indicates that the Coulomb gauge is a
convenient and reliable gauge fixing, especially in the
perturbative applications of C-S field theory.
Let us remember that, despite of the fact that the theory does not
contain degrees of freedom, the perturbative calculations
play a relevant role, for instance in the computation of knot
invariants~\cite{witten,gmm,cotta,axelrod,alr}.
Contrary to what happens in the covariant gauges,
where it becomes more and
more difficult to evaluate the radiative corrections
as the loop number increases \cite{alr,gmm,chaichen}, 
in the Coulomb gauge
only the tree level contributions to the Green
functions survive. This feature is
particularly useful in the
case of non-flat manifolds, where the momentum representation does not
exist. As an application, the
Feynman rules of C--S field theories on Riemann surfaces have been
derived in section 4.
Moreover, the analysis performed using the Dirac's formalism has shown that the
the CR's (\ref{maincomrel})
are perfectly well defined and do not lead
to ambiguities in the quantization of the
C--S models in the Coulomb gauge.
In particular, it has been verified the consistency of the CR's
(\ref{maincomrel})
with the constraints and the covariance of the theory under the
Poincar\'e group.

Despite of these positive results, there are still many open questions
concerning the use of the Coulomb gauge.
For instance we have seen that, in this gauge, the C--S theories
become two dimensional models, so that it is lecit to ask how it is possible
to compute three dimensional link invariants.
At the lowest order, where the link invariant is the simple Gauss
invariant, one can check that the results obtained in the Coulomb
gauge are consistent with those obtained in the covariant gauge (see appendix).

Another problem already mentioned is the derivation of the gauge field
propagator starting from the commutation relations (\ref{maincomrel}).

Finally one would also apply the prescription (\ref{spureg})
also to the more complicated case of the four dimensional Yang--Mills
field theories.

\section*{Acknowledgments}
The work of F. Ferrari has been supported in part by the European
Union, TMR programme, under grant ERB4001GT951315.

\section*{Appendix: Wilson loop in the abelian case}
Let us consider for instance in the abelian case
the vacuum expectation value
$$\langle W(C)\rangle = \langle e^{i\oint_C dx^\mu A_\mu}\rangle$$
for a single closed loop $C$.
At the lowest order we have:
\begin{equation}
\langle W(C)\rangle\sim -i\frac{2\pi}{s}\varphi(C)\label{wloop}
\end{equation}
where
$$\varphi(C) = \frac 1{4\pi}\oint_C dx^0\oint_C dy^i
\epsilon_{ij}
\partial^j\text{log}|{\bf x} - {\bf y}|\delta(x^0 -y^0)$$
If the loop $C$ lies on a plane, it is easy to see that the
Wilson loop (\ref{wloop}) is trivial.
If the loop is not planar, using Stokes theorem, we obtain:
\begin{equation}
\varphi(C) = \frac 1{4\pi}\oint_C dx^0\int_{\Sigma_0} d^2 y\delta^{(3)}
(x - y) \label{fc}
\end{equation}
where $\Sigma_0$ is the projection on the plane $x_1,x_2$ of a surface
$\Sigma$
spanned by the loop $C$.
This result is to be compared with what we would obtain in the
covariant gauge:
$$\varphi(C)_{cov} = \frac 1{4\pi}\oint_C dx^\mu\int_\Sigma d^2 S_\mu
\delta^{(3)}
(x - y)$$
where now $d^2 S_\mu$ is the infinitesimal area element on the surface
$\Sigma$.
Introducing a framing \cite{witten} with framing contour $C_f$
in equation (\ref{fc}) we have:
$$\varphi_f(C) = \frac 1{4\pi}\oint_{C_f} dx^0\int_{\Sigma_0} d^2 y\delta^{(3)}
(x - y)$$
After a few calculations one finds that the above integral is exactly
a
Cauchy integral counting how many times the loop $C_f$ is intersecting
the loop $C$, which is exactly the Gauss link invariant as expected.


\section*{References}
\begin{thebibliography}{99}
\bibitem{jao} R. Jackiw and S. Templeton, {\it Phys. Rev.} {\bf D23} (1981),
2291; S. Deser, R. Jackiw and S. Templeton, {\it Phys. Rev. Lett.}
{\bf 48} (1983), 975;
J. Schonfeld, {\it Nucl. Phys.} {\bf B185} (1981), 157.
\bibitem{hageno}C. R. Hagen, {\it Ann. Phys.} (NY) {\bf 157} (1984), 342.
\bibitem{csapps} G. Moore and N. Seiberg, {\it Phys. Lett.} {\bf B220} (1989),
422; J. Fr\"ohlich and C. King, {\it Comm. Math. Phys.} {\bf 126}
(1989), 167; J. M. F. Labastida and A. V. Ramallo, {\it Phys. Lett.}
{\bf 238B} (1989), 214; M. Bos and V. P. Nair,
{\it Int. Jour. Mod. Phys.} {\bf A5} (1990), 989;
W. Chen, G. W. Semenoff and Y. S. Wu, {\it Mod. Phys. Lett.} {\bf A5}
(1990), 1833 {\it Phys. Rev.} {\bf D46} (1992), 5521;
E. Guadagnini, M. Martellini and M. Mintchev, {\it Nucl. Phys.}
{\bf B336} (1990), 581; 
G. W. Semenoff, {\it Phys. Rev. Lett} {\bf 61} (1988), 517; E. Fradkin {\it
Phys. Rev. Lett.} {\bf 63} (1989), 322; M. L\"uscher, {\it Nucl. Phys.}
{\bf B326} (1989), 557; 
E. Witten, {\it Nucl. Phys.} {\bf B311} (1988), 46; 
Y. H. Chen, F. Wilczek, E. Witten and B. I. Halperin, {\it Int. Jour. Mod.
Phys.} {\bf B3} (1989), 1001;
\bibitem{ienkur}
R. Iengo and K. Lechner, {\it Phys. Rep.} {\bf 213} (1992), 179.
\bibitem{witten} E. Witten, {\it Comm. Math. Phys.} {\bf 121} (1989), 351. 
\bibitem{gmm} E. Guadagnini, M. Martellini and M. Mintchev,
{\it Phys. Lett.} {\bf B227} (1989), 111.
\bibitem{cotta} P. Cotta Ramusino, E. Guadagnini, M. Martellini and
M. Mintchev, {\it Nucl. Phys.} {\bf B330} (1990), 557.
\bibitem{axelrod} S. Axelrod and I. M. Singer, in {\it Proceedings
of the XXth International Conference on Differential Geometrical
Methods in Theoretical Physics}, New York 1991,  
edited by S. Catto and A. Rocha (World Scientific, Singapore, 1992).
\bibitem{hagent} C. R. Hagen, {\it Phys. Rev.} {\bf D31} (1985),
2135.
\bibitem{jat} S. Deser, R. Jackiw and S. Templeton,
{\it Ann. Phys.} (N. Y.) {\bf 140} (1984), 372.
\bibitem{vo}  A. Desnieres de Veigy and S. Ouvry, {\it Phys. Lett.} {\bf 387B%
} (1993), 91; D. Bak and O. Bergman, {\it Phys.
Rev.} {\bf D51} (1995), 1994; O. Bergman and G. Lozano,
{\it Ann. Phys.} (NY) {\bf 229} (1994), 229.
\bibitem{devone} F. P. Devecchi, M. Fleck, H. O. Girotti,
M. Gomes and A. J. da Silva, {\it Ann. Phys.} {\bf 242} (1995), 275.
\bibitem{devtwo} A. Foerster and H. O. Girotti, {\it Statistical transmutations
in $2+1$ dimensions}, in J. J. Giambiagi Festschrift, eds. H. Falomir, R. E.
Gamboa Saravi, P. Leal Ferreira and F. A. Schaposnik (World Scientific,
Singapore, 1990), p. 161; {\it Phys. Lett.} {\bf B230} (1989), 83;
{\it Nucl. Phys.} {\bf B342} (1990), 680.
\bibitem{bcv} A. Bellini, M. Ciafaloni and P. Valtancoli, {\it Nucl. Phys.}
{\bf B454} (1995), 449; {\bf B462} (1996), 453.
\bibitem{taylor} P. J. Doust and J. C. Taylor, {\it Phys. Lett.}
{\bf 197B} (1987), 232; P. J. Doust, {\it Ann. Phys.} (N. Y.) {\bf 177} (1987),
169.
\bibitem{chetsa} H. Cheng and E. C. Tsai, {\it Phys. Rev. Lett.}
{\bf 57} (1986), 511. 
\bibitem{leibbrandt} G. Leibbrandt, {\it Noncovariant Gauges}, World
Scientific, Singapore, 1994.
\bibitem{leiwil}
G. Leibbrandt and J. Williams,
{\it Nucl. Phys. } {\bf B475} (1996), 469 and references therein.
\bibitem{ffprd} F. Ferrari, {\it Phys. Rev.} {\bf D50} (1994), 7578.
\bibitem{cscgform} K. Haller and E. L. Lombridas, {\it Ann. Phys.} {\bf 246} (1996), 1;
M.-I. Park and Y.-J. Park, {\it Phys. Rev.} {\bf D50} (1994),
7584; S.-S. Feng, H.-S. Zong, Z.-X. Wang and X.-J. Qiu,
{\it Induced Electronic Interactions in Chern--Simons Systems},
{\it Int. Jour. Theor. Phys.}, {\bf 36} (8), (1997), 1743.
\bibitem{frig} F. Ferrari and I. Lazzizzera, 
{\it Phys. Lett.} {\bf B395}, (1997), 250.
\bibitem{flpert} F. Ferrari and I. Lazzizzera, {\it Perturbative
Analysis of the Chern--Simons Field Theory in the Coulomb Gauge},
Preprint UTF 387/96, PAR-LPTHE 96-44. {\it Int. Jour. Mod. Phys.}
{\bf A}, in press.
\bibitem{ffunp} F. Ferrari, {\it On the Quantization of the Chern--Simons
Field Theory on Curved Space--Times: the Coulomb Gauge Approach},
Preprint LMU-TPW 96-5, hep-th/9303117. 
\bibitem{linni} Q.-G. Lin and G.-J. Ni, {\it Class. Quantum Grav.} {\bf 7}
(1990), 1261.
\bibitem{covgau} L. Alvarez--Gaum\'e, J. M. F. Labastida and A. V. Ramallo,
{\it Nucl. Phys.} {\bf B334} (1990), 103;
W. Chen, G. W. Semenoff and Y. S. Wu, {\it Mod. Phys. Lett.} {\bf A5}
(1990), 1833 {\it Phys. Rev.} {\bf D46} (1992), 5521;
D. Birmingham, M. Rakowsky and G. Thompson, {\it Phys. Lett.}
{\bf B251} (1990), 121; E. Guadagnini, M. Martellini and M. Mintchev,
{\it Phys. Lett.} {\bf B227} (1989), 111;
A. Blasi and R. Collina, {\it Nucl. Phys.} {\bf B345} (1990), 472;
M. Asorey and F. Falceto, {\it Phys. Lett.} {\bf B241} (1990), 31;
F. Delduc, C. Lucchesi, O. Piguet and S. P. Sorella,
{\it Nucl. Phys.} {\bf B346} (1990), 313.
\bibitem{csformal} A. Blasi and R. Collina, {\it Nucl. Phys.} {\bf B354}
(1990), 472; F. Delduc, C. Lucchesi, O. Piguet and S. P. Sorella,
{\it Nucl. Phys.} {\bf B346} (1990), 313.
\bibitem{ss} O. Piguet and S. P. Sorella, {\it Algebraic Renormalization},
Springer Verlag 1995.
\bibitem{shift} G. Giavarini, C. P. Martin and F. Ruiz Ruiz,
{\it Phys. Lett.} {\bf B332} (1994), 345; {\it Phys. Lett.} {\bf B314}
(1993), 328;
{\it Nucl. Phys.}
{\bf B381} (1992), 222; C. P. Martin, {\it Phys. Lett.} {\bf B241} (1990),
513; M. Asorey, F. Falceto, J. L. Lopez and G. Luz\'on, {\it Phys. Rev.}
{\bf D49} (1994), 5377; M. Asorey and F. Falceto, {\it Phys. Lett.}
{\bf B241} (1990), 31.
\bibitem{alr} L. Alvarez--Gaum\'e, J. M. F. Labastida and A. V. Ramallo,
{\it Nucl. Phys.} {\bf B334} (1990), 103;
\bibitem{schwinger} J. Schwinger, {\it Phys. Rev.} {\bf 125} (1962), 1043.
\bibitem{flprep} F. Ferrari and I. Lazzizzera, work in progress.
\bibitem{itzu} C. Itzykson and J.-B. Zuber, {\it Quantum Field Theory},
McGraw--Hill, Singapore 1980.
\bibitem{empi} S. Emery and O. Piguet, {\it Helv. Phys. Acta} {\bf 64}
(1991), 1256.
\bibitem{chaichen} M. Chaichian and W. F. Shen, {\it Two loop
Finiteness of Chern--Simons Field Theory in Background Field
Method}, hep-th/9607208.
\bibitem{gribov}
V. N. Gribov, {\it Nucl. Phys.} {\bf B139} (1978),1;
P. van Baal, {\it Nucl. Phys.} {\bf 369} (1992), 259.
\bibitem{vv} E. Verlinde, H. Verlinde, {\it Nucl. Phys.} {\bf B288}
(1987), 357.







\end{thebibliography}
\end{document}